\documentclass[reprint,prl,superscriptaddres,nofootinbib,superscriptaddress]{revtex4-1}
\usepackage{bm}

\newcommand{\aap}{{Astron.~Astrophys.}}
\newcommand{\jcap}{{JCAP}}
\newcommand{\apjl}{{Astrophys.~J.~Lett.}}
\newcommand{\apjs}{{Astrophys.~J.~Supp.}}
\newcommand{\aj}{{Astron.~J.}}
\newcommand{\mnras}{{Mon.~Not.~R.~Astron.~Soc.}}

\newcommand{\ra}[1]{\renewcommand{\arraystretch}{#1}}

\newcommand{\lcdm}{$\Lambda$CDM}
\newcommand{\mnu}{\sum m_{\nu}}
\newcommand{\ev}{\, \rm{eV}}

\newcommand{\LCDM}{$\Lambda\rm{CDM}$}

\usepackage{graphicx}
\usepackage[large]{subfigure}
\usepackage{amssymb, amsmath}
\usepackage[amssymb]{SIunits}
\usepackage{aas_macros}
\usepackage{natbib}
\renewcommand\section[1]{\emph{#1}.---}

\usepackage{xcolor}
\usepackage{booktabs}
\usepackage{multirow} 
 
\begin{document}

\title{Cosmological limits on the neutrino mass sum for beyond-$\Lambda$CDM models}

\author{Helen~Shao}\email{hshao@alumni.princeton.edu}
\affiliation{Department of Astrophysical Sciences, Peyton Hall, Princeton University, Princeton, NJ 08544, USA}

\author{Jahmour J. Givans}
\affiliation{%
 Center for Computational Astrophysics, Flatiron Institute, 162 5th Ave, New York, NY 10010, USA
}%
\affiliation{Department of Astrophysical Sciences, Peyton Hall, Princeton University, Princeton, NJ 08544, USA}

\author{Jo Dunkley}
\affiliation{Department of Physics, Jadwin Hall, Princeton University, Princeton, NJ 08544, USA}
\affiliation{Department of Astrophysical Sciences, Peyton Hall, Princeton University, Princeton, NJ 08544, USA}

\author{Mathew Madhavacheril}
\affiliation{Department of Physics and Astronomy, University of Pennsylvania, 209 South 33rd Street, Philadelphia, PA, USA 19104}

\author{Frank Qu} 
\affiliation{DAMTP, Centre for Mathematical Sciences, University of Cambridge, Wilberforce Road, Cambridge CB3 OWA, UK}\affiliation{Kavli Institute for Cosmology Cambridge, Madingley Road, Cambridge CB3 0HA, UK}

\author{Gerrit Farren}
\affiliation{DAMTP, Centre for Mathematical Sciences, University of Cambridge, Wilberforce Road, Cambridge CB3 OWA, UK}\affiliation{Kavli Institute for Cosmology Cambridge, Madingley Road, Cambridge CB3 0HA, UK} 

\author{Blake Sherwin}
\affiliation{DAMTP, Centre for Mathematical Sciences, University of Cambridge, Wilberforce Road, Cambridge CB3 OWA, UK}\affiliation{Kavli Institute for Cosmology Cambridge, Madingley Road, Cambridge CB3 0HA, UK}

\date{\today}
\smallskip

\begin{abstract}
The sum of cosmic neutrino masses can be measured cosmologically, as the sub-eV particles behave as `hot' dark matter whose main effect is to suppress the clustering of matter compared to a universe with the same amount of purely cold dark matter. Current astronomical data provide an upper limit on $\mnu$ between 0.07 - 0.12~eV at 95\% confidence, depending on the choice of data. This bound assumes that the cosmological model is $\Lambda$CDM, where dark energy is a cosmological constant, the spatial geometry is flat, and the primordial fluctuations follow a pure power-law. Here, we update studies on how the mass limit degrades if we relax these assumptions. To existing data from the \textit{Planck} satellite we add new gravitational lensing data from the Atacama Cosmology Telescope, the new Type Ia Supernova sample from the Pantheon+ survey, and baryonic acoustic oscillation (BAO) measurements from the Sloan Digital Sky Survey and the Dark Energy Spectrosopic Instrument. We find the neutrino mass limit is stable to most model extensions, with such extensions degrading the limit by less than 10\%. We find a broadest bound of $\mnu < 0.19 ~\rm{eV}$ at 95\% confidence for a model with dynamical dark energy, although this scenario is not statistically preferred over the simpler \lcdm\ model.

\end{abstract}
\maketitle

\section{Introduction}
Neutrinos are electrically uncharged spin-$1/2$ fermions which exist in one of three active flavor states: electron, muon and tau neutrinos ($\nu_e$, $\nu_\mu$, $\nu_\tau$). The discovery of neutrino flavor oscillations \cite{SuperK_1998,SNO_2001,SNO_2002} showed that neutrinos have mass, with each neutrino flavor occupying a superposition of three mass eigenstates, $m_i$ ($i = 1,2,3$). Oscillation experiments measure the mass-squared splittings between these states, $\Delta m_{ij}^2 = m_i^2 - m_j^2$, with the mass states either in an inverted hierarchy (IH), where $m_3\ll m_1<m_2$, or the normal hierarchy (NH), where $m_1<m_2\ll m_3$. In the normal hierarchy the sum of the masses, $\sum m_i$, has a lower limit of $0.06$~eV; in the inverted hierarchy it is $0.1$~eV \cite{2019JHEP...01..106E}. 

Direct-detection $\beta$-decay experiments  give limits of $m \leq 0.8 \ev$ (90\% confidence level) for the electron neutrino by the KATRIN Collaboration \citep{aker2021direct}.
Future experiments are  expected to reach a sensitivity of $0.2 \, \rm{eV}$ (90\% c.l.) \cite{KATRIN_report}. A tighter indirect limit has already been placed on the sum of neutrino masses from cosmological data \citep{PhysRevD.70.045016, Lesgourgues_2006, Jimenez_2010, Hamann_2012, Hannestad_2016, Vagnozzi_2018,planck_pr3,Choudhury_2020, wang2024}, with the primary effect on cosmology being the suppression of the clustering of matter compared to a universe with the same amount of purely cold dark matter. Expressed as the sum of neutrino masses, the current upper bound at 95\% confidence is in the range $\mnu < 0.07-0.12 ~\rm{eV}$ depending on the choice of data, with the tightest limit obtained when including new baryon acoustic oscillation (BAO) data from the Dark Energy Spectroscopic Instrument (DESI) \cite{Anderson_2012,planck18,act_dr6,pr4_desi_sdss,desi}.
This bound assumes that the correct cosmological model is \lcdm, where dark energy is a cosmological constant, the spatial geometry is flat, cold dark matter interacts only via gravity, and the power spectrum of primordial perturbations follows a pure power-law. Previous studies have examined how the mass limit degrades if we loosen these assumptions, since some model extensions can mimic the effect of non-zero neutrino mass on the clustering of matter \citep[e.g,][]{Allison_2015,Choudhury_2018,Choudhury_2020, pr4_desi_sdss,desi,elbers2024,naredotuero2024}. In this paper we update bounds using the latest data from the Atacama Cosmology Telescope (ACT) \cite{qu2023atacama,act_dr6}, the Pantheon+ supernova survey \cite{pantheonp}, and a composite of BAO data from DESI and the Sloan Digital Sky Survey, allowing for variations in, for example, the spatial curvature and the equation of state of dark energy. 
\begin{figure}
    \centering
    \includegraphics[width=\linewidth]{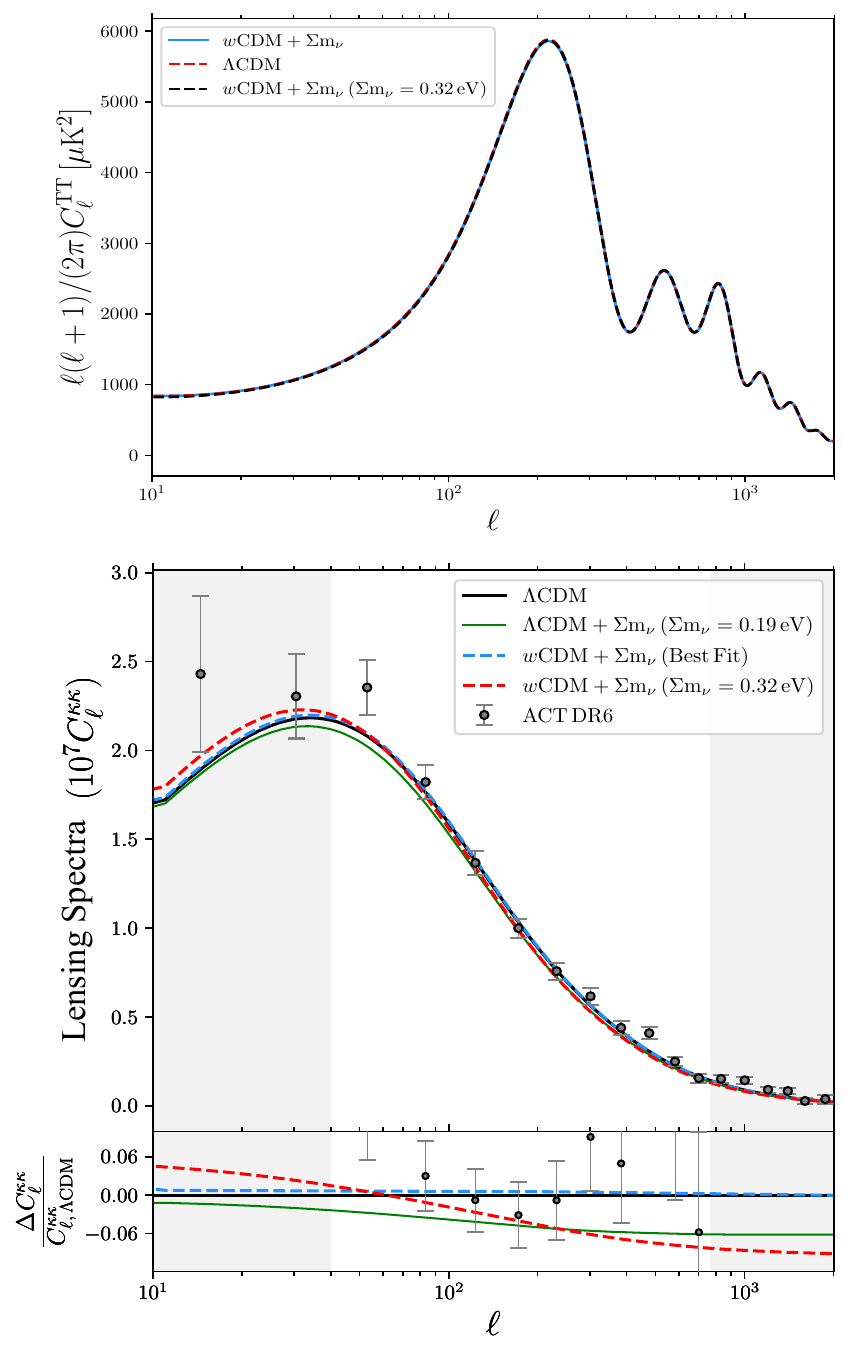}
    \caption{A \LCDM\ model with a neutrino mass sum of $0.19$~eV (green), and a CDM model with dark energy equation of state $w$ and neutrino mass sum of $0.32$~eV (red), are not distinguishable from a zero neutrino mass model using the CMB intensity power spectrum (top). However, neutrino mass sums of this size, within these different scenarios, can both be excluded at 3$\sigma$ significance using current data combination. The lower panel shows that their CMB lensing power spectra can be $\sim$6-9\% lower than for the zero-mass \LCDM\ model, and are disfavored by the indicated ACT DR6 lensing data.}
    \label{fig:spectra}
\end{figure}
\section{Methodology}\label{sec:methods}
We follow similar analysis methods as in \cite{Choudhury_2020,act_dr6}, estimating the posterior distribution for sets of cosmological parameters that include a varying neutrino mass sum, using a mixture of cosmological datasets.  This includes the angular power spectra of the CMB intensity and polarization anisotropy as measured by the \textit{Planck} satellite \cite{planck18,rosenberg22}. We note that the PR3 \textit{Planck} likelihood showed a 2.9$\sigma$ inconsistency between \lcdm\ predictions of the CMB lensing parameter $A_L$ and its value as measured via its smearing effect on the temperature and polarization power spectra \cite{planck_pr3}. This led to artificially tightened constraints on the neutrino mass. Hence, for our baseline data, as in \cite{act_dr6}, we choose to use the \textit{Planck} likelihood constructed from the PR4 NPIPE processed maps \cite{rosenberg22}. We also include BAO measurements which characterize the typical separation of galaxies and provide a measure of the total amount of dark matter. Our `BAO-1' dataset includes data measured by the 6dF \citep{Beutler_2016} and Sloan Digital Sky Survey DR7 \citep{Percival_2010}, the latter of which is combined with the BOSS DR12 and DR16 data \citep{BOSS_bao}. We also use a composite `BAO-2' dataset combining results from both SDSS and DESI,  described in the appendix.\footnote{\raggedright We use the following \texttt{Cobaya v3.5.1} \citep{Torrado_2021} likelihoods for primary CMB data: \texttt{planck\_2018\_lowl.TT, planck\_2018\_lowl.EE\_sroll2, planck\_NPIPE\_highl\_CamSpec.TTTEEE}, and for BAO: \texttt{bao.sixdf\_2011\_bao, bao.sdss\_dr7\_mg, bao.sdss\_dr16\_baoplus\_lrg}.}

To this baseline data combination, we add the angular power spectrum of the reconstructed gravitational lensing of the CMB measured by \textit{Planck} \cite{pr4_lens} and ACT \citep{qu2023atacama}; this provides a measure of the growth of cosmic structure at later times, peaking at typically half the age of the universe. We label these datasets as: `CMB (Planck)' and `CMB (ACT+\textit{Planck})'. For the latter, we emphasize that we are using a variation of the ACT likelihood that includes \textit{Planck} lensing data. Finally, we include the distances to Type Ia supernova as measured by the Pantheon+ survey \cite{pantheonp}, offering a direct way to measure the expansion rate of the universe. Here, we do not include the SH0Es calibration \citep{Riess_2022} and refer to this dataset as `SNe'.\footnote{\raggedright We convert the Pantheon+ likelihood used with CosmoSIS to a version compatible with Cobaya and check we get the same results as using \citep{bidenko}.} Consistent constraints on $\mnu$ have also been placed using the full-shape power spectrum of galaxies and redshift space distortions (e.g., \cite{Upadhye_2019}), though we do not use these in our analyses. 

Our baseline model is \lcdm, characterized by two initial conditions (the amplitude and spectral index of power-law scalar fluctuations, $A_s$ and $n_s$, respectively), three ingredients (baryon density, cold dark matter density and dark energy fraction), and an optical depth to reionization, $\tau$, to which we add the sum of neutrino masses. The extension model parameters we consider are the curvature, $\Omega_k$, the equation of state of dark energy that is either constant with time or varies with time, $w(z)$, the effective number of neutrino species, $N_{\rm eff}$, and a running of the primordial scalar spectral index, $n_\mathrm{run}$\footnote{\raggedright In \lcdm\ there are 3.046 effective neutrino species and a power law spectral index with no running.}. As in \cite{act_dr6}, we generate theoretical predictions for these models with the CAMB numerical code, using the Parameterized Post-Friedmann (PPF) prescription for dark energy perturbations \citep{ppf}, and the Chevallier-Polarski-Linder (CPL) parameterization of the form \citep{2001IJMPD..10..213C,2003PhRvL..90i1301L}, 
\begin{equation}\label{eqn:CPL}
    w(z) = w_0 + w_a\left(\frac{z}{1+z}\right)
\end{equation}
when we consider variations of dark energy equation of state as a function of redshift.\footnote{\raggedright To compute the small-scale nonlinear matter power spectrum we use the \textsc{halofit} model \citep{smith} implemented with HMcode \citep{Takahashi_2012} and the \textsc{MEAD} fitting method \citep{mead_2015}.} We estimate the posterior distribution of parameters with Metropolis Hastings methods using the \texttt{Cobaya} sampling code, quoting marginalised one-dimensional medians or 95\% upper limits. To remain independent of neutrino oscillation experiments, we do not impose the lower mass limit for either NH or IH as a prior. However, we do require $\mnu$ to be positive. Previous works have also employed hierarchy-agnostic priors \citep[e.g][]{Mahony_2020, Choudhury_2020} but did not find conclusive results on a preferred hierarchy. 

As described in \citep{Allison_2015}, in this baseline model the neutrino mass is not strongly correlated with any of the other six parameters, but it has weak correlations with the primordial amplitude and the optical depth to reionization as they both affect the amplitude of the clustering. Fig. \ref{fig:spectra} shows how a model with $\mnu=0.19$~eV cannot be distinguished from a zero neutrino mass model using only primary CMB anisotropy data---neutrinos would have become non-relativistic only after the CMB formed. However, such a model would lower the CMB lensing signal by about 6\%, which is sufficient to be disfavored at the 3$\sigma$ significance level. 
\begin{figure}
    \centering
    \includegraphics[width=.99\linewidth]{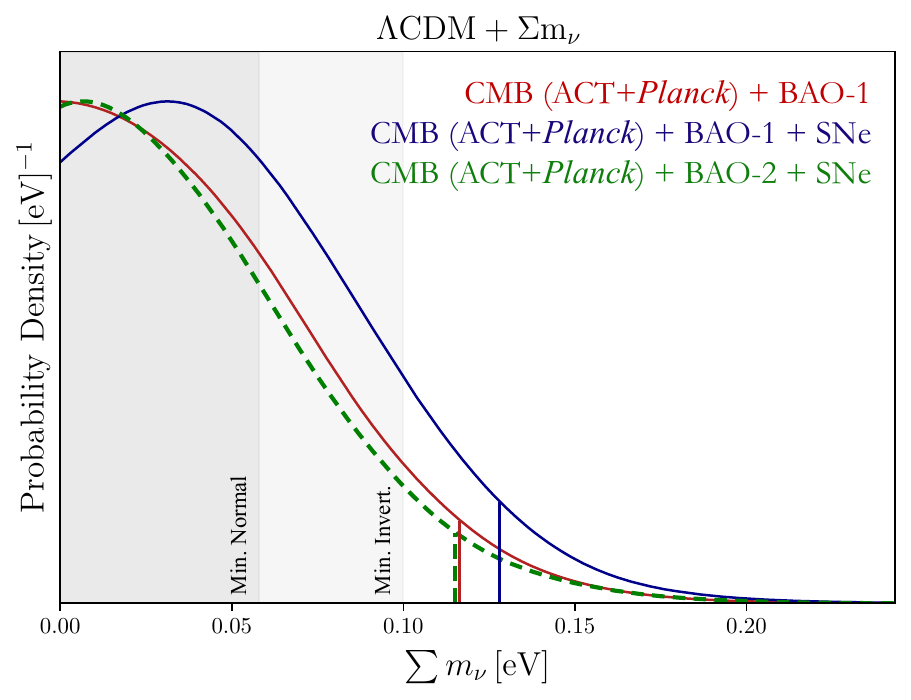}
    \vspace{-0.3cm}
    \caption{The 1D marginalized posterior distributions for $\mnu$ in the \lcdm+$\mnu$ model, with the lower bounds on neutrino mass sum required for the normal ($\mnu < 0.06$ eV) and inverted ($\mnu < 0.1$ eV) hierarchies shaded. The vertical lines indicate the 95\% upper bound for each dataset. All three datasets include primary CMB data from \textit{Planck} and CMB lensing from  ACT and \textit{Planck}. 'BAO-1' refers to SDSS measurements while 'BAO-2' refers to a mixture of SDSS+DESI data, as described in the text. Including SNe data results in a non-zero peak in $\mnu$.
    }
    \label{fig:base_mnu_1d}
\end{figure}

\section{Results} We initially obtain similar results to \cite{act_dr6}, finding $\mnu < 0.12 \, \rm{eV}$ for the \lcdm\ model, before including SNe data.\footnote{\raggedright \cite{act_dr6} uses different BAO likelihoods, resulting in slightly different $\mnu$ constraints.} Including the SNe data shrinks the $1\sigma$ uncertainty by a small amount, but the preferred value is slightly non-zero, so the upper limit increases to $0.13 \, \rm{eV}$, as shown in Fig. \ref{fig:base_mnu_1d}. If we incorporate DESI measurements, using the CMB (ACT+\textit{Planck}) + BAO-2 + SNe dataset, the constraint marginally tightens to $\mnu<0.12$~eV, but we still find a non-zero peak in the $\mnu$ posterior distribution as shown in Fig.~\ref{fig:base_mnu_1d}. This is a higher upper bound compared to the limit reported in \citep{desi}, since our BAO-2 dataset conservatively does not include the new DESI data in the first two redshift bins, using SDSS data instead. With this choice of data the inverted hierarchy is still viable in the minimal extension. This is consistent with the findings of \citep{pr4_desi_sdss, naredotuero2024}, who provide analogous constraints using the \texttt{HiLLiPop+LoLLiPoP} \textit{Planck} likelihoods \citep{Tristram_2024} and a slightly different combination of DESI+SDSS data. A summary of the 95\% upper limits for different datasets is given in Table \ref{tab4}.
\begin{figure}[!]
    \centering
    \includegraphics[width=.99\linewidth]{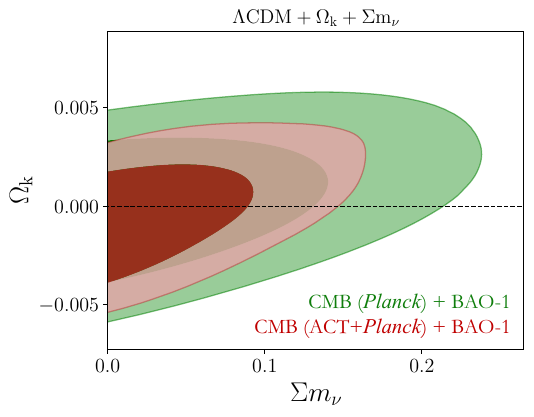}
    \vspace{-0.3cm}
    \caption{2D marginalized distribution for $\Omega_k$ and $\mnu$ in the $\Lambda\rm{CDM} + \Omega_{\rm k} + \mnu$ cosmology. A spatially flat universe has $\Omega_k=0$, an open universe has $\Omega_k>0$, and a closed universe has $\Omega_k<0$. Including the new ACT CMB lensing data gives a tighter constraint than with earlier lensing data from Planck. Adding BAO-2 or SNe data does not further constrain $\mnu$.
    \label{fig:mnu_k}}
\end{figure}
\begin{figure}[!]
    \centering
    \includegraphics[width=0.98\linewidth]{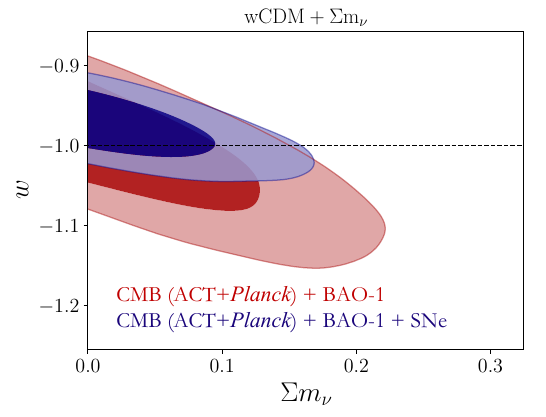}
    \vspace{-0.3cm}
    \caption{The 2D marginalized posterior for $w$ and $\mnu$ in the $w\rm{CDM}+\mnu$ cosmology, where $w=-1$ is a cosmological constant.  Adding SNe data reduces the degeneracy between $w$ and $\mnu$, resulting in $\mnu <0.13 \, \rm{eV}$, similar to the baseline model.
    \label{fig:mnu_w}}
\end{figure}
In Fig. \ref{fig:mnu_k} we additionally vary the spatial curvature, where $\Omega_k<0$ ($>0$) corresponds to a closed (open) universe. As has been shown in e.g., \cite{Allison_2015,planck18,Choudhury_2018,Choudhury_2020},
a slightly larger neutrino mass is allowed in an open universe with $\Omega_k>0$. Due to the geometric degeneracy \citep{Efstathiou_1999}, an open universe needs more dark energy to conserve the peak positions in the primary CMB spectrum.

These models would have a smaller CMB lensing signal than the best-fitting \lcdm\ model \citep{Sherwin_2011}, which can be partly compensated by decreasing the neutrino mass. However, BAO data provides orthogonal constraints on $\mnu$ and curvature, switching the direction of this degeneracy \cite{Allison_2015}. The new ACT lensing data better measures the clustering, resulting in $\mnu<0.13$ eV compared to 0.17~eV with only \textit{Planck} lensing.

Including the SNe data does not further tighten constraints for this model. 
\begin{figure}
    \centering
    \includegraphics[width=0.95\linewidth]{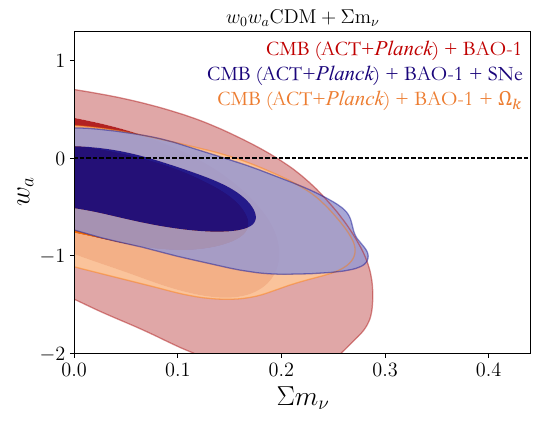}
    \includegraphics[width=0.95\linewidth]{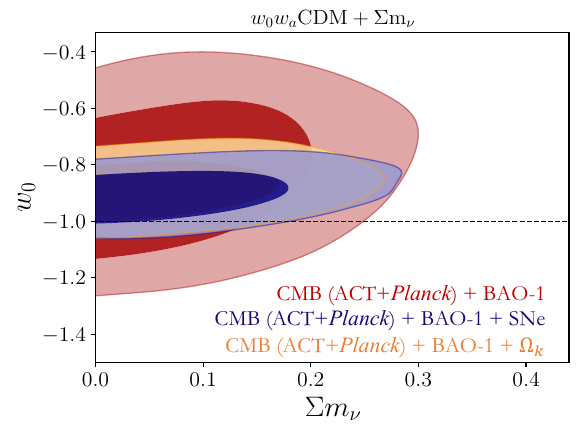}
    \includegraphics[width=0.95\linewidth]{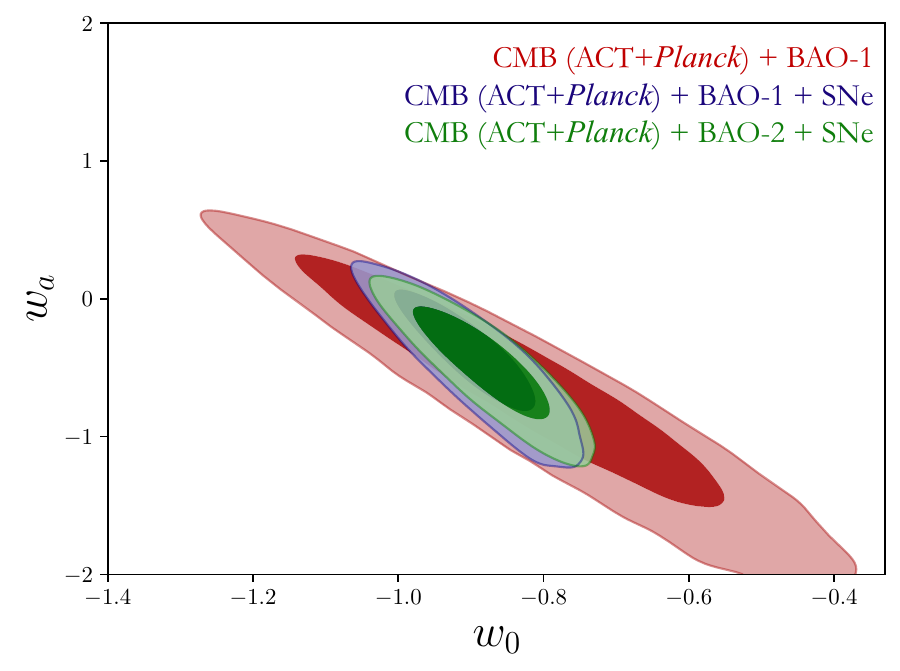}
    \caption{The 2D marginalized posteriors for the $w_aw_0\rm{CDM}+\mnu$ model where the equation of state is a function of $z$ as in Eqn.~\ref{eqn:CPL}. The indicated $w_0=-1$ and $w_a=0$ correspond to a cosmological constant. Adding SNe data significantly tightens the constraints on $w_a$ and $w_0$, but has only a modest effect on the $\mnu$ bound. Including BAO-2 measurements, indicated by the green contours, do not affect the dark energy constraints but reduces the tail on $\mnu$ to $<0.19$ eV (95\% c.l), as seen in the second plot. Moreover, additionally varying $\Omega_{\rm k}$ does not affect the constraint, as demonstrated by the orange contour in the first plot.}
    \label{fig:2d_w0waCDM}
\end{figure}
\begin{figure*}[!]
    \centering
    \includegraphics[width=1.0\linewidth]{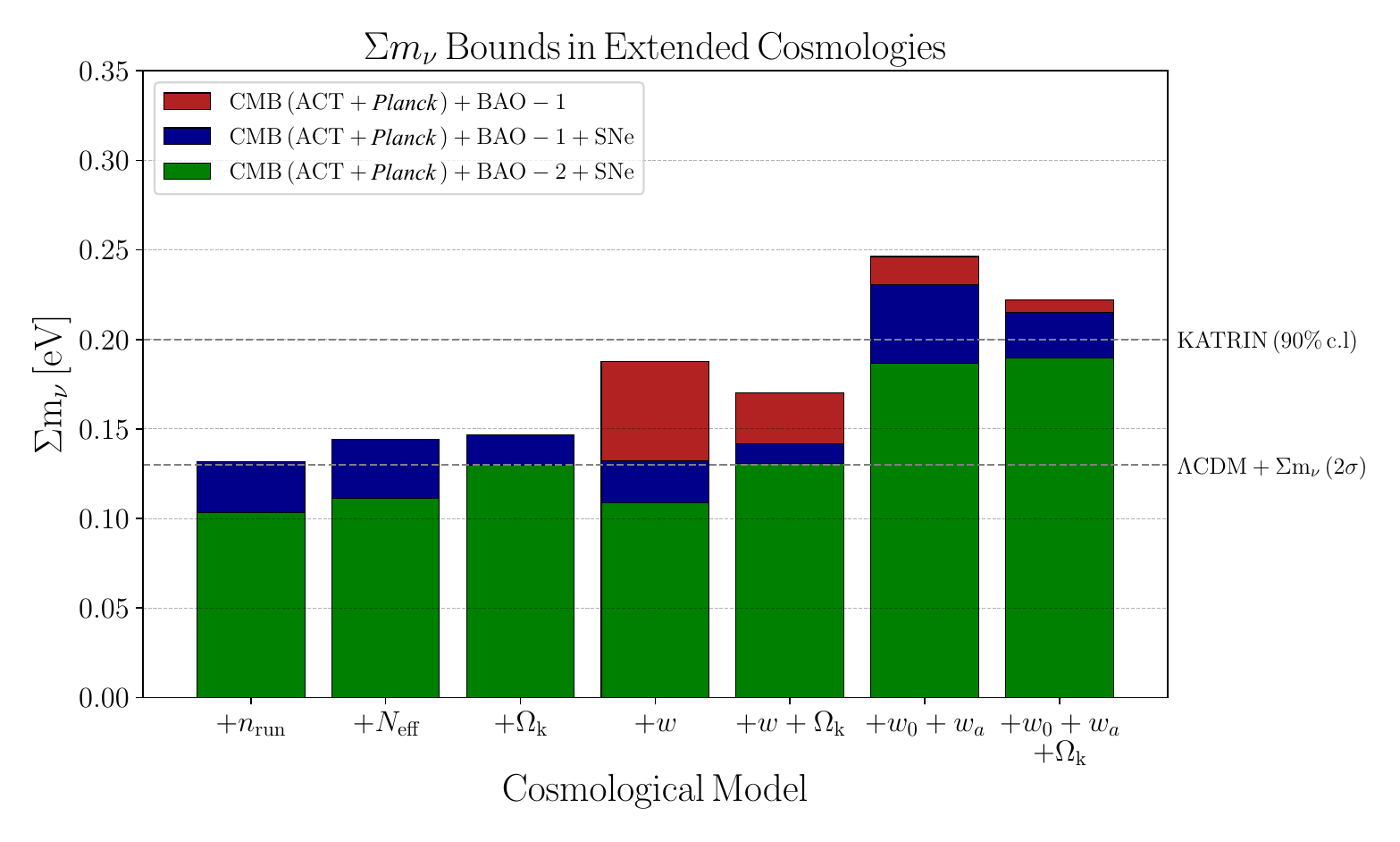}
     \caption{Summary of upper limits, at 95\% confidence, on the sum of neutrino masses for the extended models indicated on the horizontal axis. The $\mnu$ bound obtained by the blue dataset for the \LCDM+$\mnu$ model is indicated at $0.13 ~\rm{eV}$, and the line at $0.2$ eV denotes the expected future bound from KATRIN on the electron neutrino mass \citep{aker2021direct}. As expected, the effect of adding SNe data is most apparent for the extensions with dark energy. Including BAO-2 measurements tightens the constraints for all models, achieving an upper bound of $\mnu<0.11~\rm{eV}$ in the minimally extended cosmology.}
    \label{fig:histograms}
\end{figure*}
If the dark energy is not a constant vacuum energy, its equation of state, $w=p/\rho$, is anti-correlated with neutrino mass. As discussed in \cite{Hannestad_2005,Choudhury_2020}, if $w<-1$, the dark energy density $\Omega_{\rm DE}(z)$ is decreased, then the total matter density must increase to fit the primary CMB data. This enhances the growth of cosmic structures, whose effects on CMB lensing can be counteracted by increasing the neutrino mass. This correlation increases the neutrino mass bound to $\mnu<0.18$~eV with CMB(ACT+\textit{Planck})+BAO-1 data. In this case, the SNe data measure the time dependence of the expansion rate of the universe, providing an additional constraint on $w$. This partly breaks the degeneracy, as shown in Fig. \ref{fig:mnu_w}, bringing the limit back to $\mnu<0.13$~eV. Including BAO-2 data further decreases this bound to $\mnu<0.11$~eV but does not improve the constraint on $w$. Moreover, allowing both $w$ and $\Omega_k$ to vary together does not further decrease the neutrino mass uncertainty. Neither $w\ne-1$ nor $\Omega_k\ne0$ are preferred by the data.

The effects of non-zero neutrino mass can only be partially mimicked by a constant equation of state of dark energy. If this component is dynamic, corresponding to a new scalar field for example, the equation of state would have a time dependence. For the time-varying model we consider, the dark energy density scales as \citep{Linder_2003}
\begin{equation}\label{eq:dde}
    \Omega_{\rm DE}(z) = \Omega_{\rm DE}(0) (1+z)^{3(1+w_0+w_a)} \textrm{exp}\left(-3w_a\frac{z}{1+z}\right).
\end{equation} 
Fig. \ref{fig:2d_w0waCDM} shows how the correlation of the neutrino mass with the equation of state today, $w_0$, is reduced, and instead becomes anti-correlated with $w_a$. This happens because $w_a$ governs the late time dynamics of dark energy, as seen in Eq.~\ref{eq:dde} for small $z$. Consequently, it has a larger effect on $H(z)$ and on $\mnu$ than does $w_0$ \citep{Choudhury_2020}. As discussed earlier, decreasing the dark energy equation of state through an increase in $w_a$ can result in a good fit to the data (namely, the measured comoving distance to decoupling) if $\mnu$ is increased. Meanwhile, the $w_0 + w_a$ term in the exponent of Eq.~\ref{eq:dde} induces anti-correlations between the two parameters, as seen in the third plot of Fig.~\ref{fig:2d_w0waCDM}. Thus, $w_0$ and $\mnu$ are now slightly correlated due to their mutual anti-correlations with $w_a$. Including the SNe data, the bound is $\mnu<0.23$~eV; without the ACT and SNe data this limit is 25\% weaker at $0.29$~eV. Using the BAO-2 data also shrinks the parameter space for larger $\mnu$, as shown the green contours of Fig.~\ref{fig:2d_w0waCDM}, resulting in $\mnu < 0.19$~eV. Finally, we find that allowing curvature to be additionally varied in this model does not further increase the uncertainty, as seen in Fig.~\ref{fig:2d_w0waCDM}.

\section{Conclusions}
With an upper limit from cosmological data on the neutrino mass sum of $\mnu<0.12$~eV, rising to $\mnu<0.19$~eV in a time-varying dark energy model, we are approaching the minimum predicted level for either neutrino mass hierarchy (0.06 or 0.1~eV). Our analysis shows that the upper limit does not depend strongly on assumptions about the cosmological model, and that the \lcdm\ model is still preferred when applying Bayesian model selection. 
A lower bound on $\mnu$ has not yet been seen with cosmological data: we have yet to find evidence for non-zero neutrino mass. Improved data from future DESI releases and the Simons Observatory, as well as measurements of the optical depth from CLASS, TauRUS or LiteBIRD, hold promise for a non-zero mass detection \citep{dvorkin2020}. In such a scenario, it will be important to distinguish a non-zero mass from a time-varying dark energy. We can expect dark energy properties will be better constrained by future supernova, weak lensing and galaxy clustering data from, e.g., the Vera C. Rubin Observatory and the Euclid satellite. Combinations of multiple cosmological datasets will be crucial for solidifying our understanding of neutrino properties beyond the Standard Model.

\begin{table*}
\centering
\ra{1.1} 
\resizebox{0.8\textwidth}{!}{
\begin{tabular}{|c|cc|cc|cc|}
\hline
\multirow{2}{*}{\textbf{Model (+$\mnu$)}}  & \multicolumn{2}{c|}{\multirow{2}{9em}{CMB (\textit{Planck}) + BAO-1 }} & \multicolumn{2}{c|}{\multirow{2}{10em}{CMB (ACT+\textit{Planck}) + BAO-1}} & \multicolumn{2}{c|}{\multirow{2}{10em}{CMB  (ACT+\textit{Planck}) + BAO-2 }} \\
& & & & & & \\
\cline{2-7}
& without SNe & with SNe & without SNe & with SNe & without SNe & with SNe \\ 
\midrule
\hline
$\Lambda \textrm{CDM}$ &$<0.14$&  $<0.15$ & $<0.12$& $<0.13$& $<0.10$ & $<0.12$ \\
\midrule
$\Lambda\textrm{CDM}+n_{\rm run}$ &$<0.13$&  $<0.14$& $<0.12$& $<0.13$ & $<0.10$ & $<0.11$ \\
$\Lambda\textrm{CDM}+N_{\rm eff}$ &$<0.13$&  $<0.15 $ & $<0.13$& $<0.14$& $<0.11$ & $<0.12$ \\
\midrule
$\Lambda\textrm{CDM}+\Omega_{\rm k}$ &$<0.17$&  $<0.15$ & $<0.13$& $<0.14$& $<0.13$ & $<0.14$ \\
\midrule
$w\textrm{CDM}$ &$<0.21$&  $<0.14$& $<0.18$& $<0.13$ & - & $<0.11$\\
\midrule
$w\textrm{CDM}+\Omega_{\rm k}$ &$<0.21$&  $<0.17$ & $<0.17$& $<0.14$& - & $<0.13$ \\
\midrule
$w_0w_a\textrm{CDM}$ &$<0.28$&  $<0.26$&$<0.25$& $<0.23$& - & $<0.19$\\
\midrule
$w_0w_a\textrm{CDM}+\Omega_{\rm k}$ &$<0.27$&  $<0.27$& $<0.23$ & $<0.22$& - & $<0.19$ \\
\hline
\bottomrule
\end{tabular}}
\caption{The 95\% upper limits on $\mnu$ $[\rm{eV}]$ in the extended cosmological models. All results are obtained using \textit{Planck} primary CMB data, and the six cases show the effects of adding ACT lensing data, DESI measurements, and/or Type Ia supernova data from Pantheon+. BAO-1 indicates SDSS-only data while BAO-2 denotes our DESI+SDSS combination. The rightmost column indicates the most constraining combination.}
\label{tab4}
\end{table*}

\section{Acknowledgments}
JG acknowledges support from Princeton's Presidential Postdoctoral Research Fellowship. JD acknowledges NSF grant AST2108126. MM acknowledges support from NSF grants AST-2307727 and  AST-2153201 and NASA grant 21-ATP21-0145.  We use the \texttt{Cobaya code}.

\section{Appendix}\label{sec:desi_sdss}
\begin{figure}
    \centering
    \includegraphics[width=1.0\linewidth]{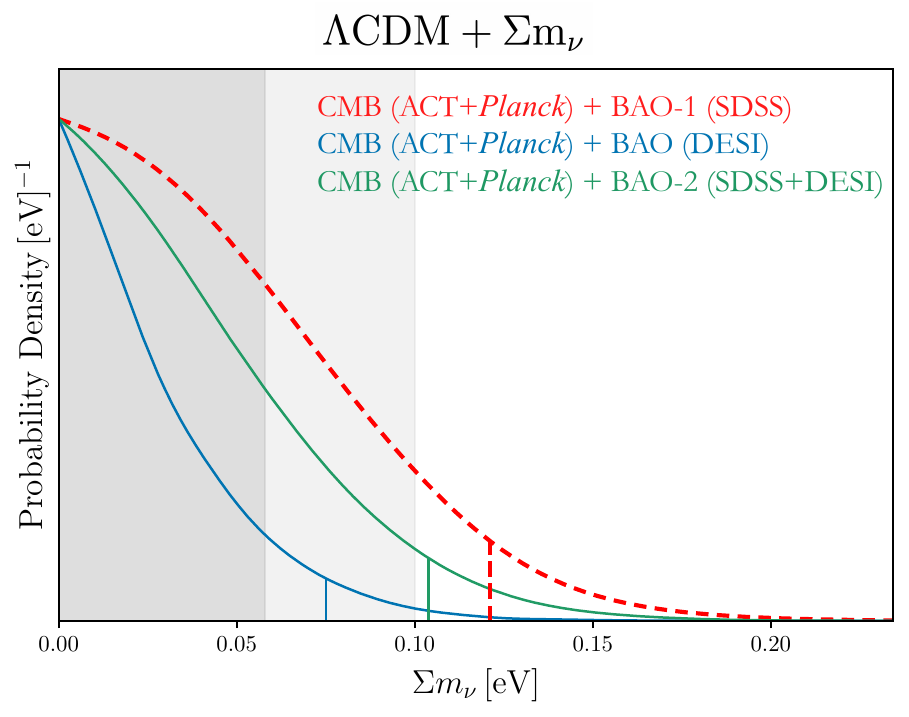}
    \caption{The 1D posterior distributions for $\mnu$ in the \lcdm+$\mnu$ model, similar to Fig.~\ref{fig:base_mnu_1d}, using different combinations of BAO measurements from SDSS and DESI as described in the Appendix. Our chosen `BAO-2' data combination (green), provides a more conservative upper bound of $\mnu < 0.104~$eV (95\% c.l) than using solely DESI measurements.}
    \label{fig:SDSS_DESI}
\end{figure}

Recent BAO measurements by DESI placed tight constraints on the upper bound of neutrino mass sum in various cosmologies \citep{desi}. Notably, they obtain $\mnu < 0.072$ eV in the \lcdm$+ \mnu$ model. Given that this bound appears to exclude the IH at 95\% confidence, we investigate the impact of this constraint when using select BAO measurements from SDSS in place of DESI. In particular, we note the $\sim 3\sigma$ difference found between the DESI and SDSS LRG results in the redshift interval $0.6 < z < 0.8$ \citep{desi_galaxies}. Thus, we choose to use a composite `BAO-2' dataset containing measurements from both DESI and SDSS, conservatively swapping in the SDSS LRG measurements for these ranges in our data mixture. A similar data combination is tested in \citep{desi, pr4_desi_sdss}, where BAO results from each redshift range are selected from the survey that covers the larger effective volume at that redshift. Our `BAO-2' consists of the following BAO measurements:
\begin{enumerate}
    \item We use the SDSS measurements at $z = 0.15$, 0.38, and 0.51 in place of the BGS and LRG1 results from DESI. The corresponding \texttt{Cobaya} likelihoods are \texttt{bao.sdss\_dr7\_mgs, sdss\_dr12\_consensus\_bao}. This coincides with the selection described in \cite{desi}, as the SDSS survey contains larger effective volume at these redshifts. 
    \item We use the SDSS measurements for $0.6 < z <  0.8$ in place of the LRG2 results from DESI, corresponding to the \texttt{sdss\_dr12\_consensus\_bao} \texttt{Cobaya} likelihood. This differs from the selection described in \cite{desi}, which used the DESI measurements to maximize effective survey volume.
    \item We use the DESI LRG+ELG combination over $0.8 < z < 1.1$, and the higher redshift ELGs and QSOs. The corresponding \texttt{Cobaya} likelihoods are \texttt{bao.desi\_2024\_bao\_lrgpluselg\_z1, bao.desi\_2024\_bao\_elg\_z2, bao.desi\_2024\_bao\_qso\_z1}.
    \item For the Ly$\alpha$ measurements, we use the combined DESI+SDSS data provided by the \texttt{bao.desi\_2024\_eboss\_bao\_lya} likelihood.
\end{enumerate}
Using this in our CMB (ACT+\textit{Planck}) + BAO-2 dataset for the minimal cosmological extension: \lcdm$ + \mnu$, we find an upper bound of $\mnu < 0.104$ eV (95\% c.l). The marginalized posterior distribution is shown in green in Fig.~\ref{fig:SDSS_DESI}. If instead we swap out the SDSS LRG measurement at $z=0.6$ with the corresponding data point from DESI, we find $\mnu < 0.081$ eV (95\% c.l). These results can be compared to the upper bound of $\mnu < 0.075$ eV (95\% c.l) obtained using CMB (ACT+\textit{Planck})+ BAO(DESI), where BAO measurements are from DESI only (blue curve in Fig.~\ref{fig:SDSS_DESI}). Hence, we find an increased ($\sim 44\%$) neutrino mass bound when excluding both DESI LRG measurements. While these upper bounds still place pressure on the IH, they allow for NH as a viable mass paradigm in this model. The constraint is further relaxed when including SNe data, as shown in Table~\ref{tab4}, in agreement with \citep{pr4_desi_sdss}.

\bibliography{refs}

\begin{thebibliography}{49}%
\makeatletter
\providecommand \@ifxundefined [1]{%
 \@ifx{#1\undefined}
}%
\providecommand \@ifnum [1]{%
 \ifnum #1\expandafter \@firstoftwo
 \else \expandafter \@secondoftwo
 \fi
}%
\providecommand \@ifx [1]{%
 \ifx #1\expandafter \@firstoftwo
 \else \expandafter \@secondoftwo
 \fi
}%
\providecommand \natexlab [1]{#1}%
\providecommand \enquote  [1]{``#1''}%
\providecommand \bibnamefont  [1]{#1}%
\providecommand \bibfnamefont [1]{#1}%
\providecommand \citenamefont [1]{#1}%
\providecommand \href@noop [0]{\@secondoftwo}%
\providecommand \href [0]{\begingroup \@sanitize@url \@href}%
\providecommand \@href[1]{\@@startlink{#1}\@@href}%
\providecommand \@@href[1]{\endgroup#1\@@endlink}%
\providecommand \@sanitize@url [0]{\catcode `\\12\catcode `\$12\catcode
  `\&12\catcode `\#12\catcode `\^12\catcode `\_12\catcode `\%12\relax}%
\providecommand \@@startlink[1]{}%
\providecommand \@@endlink[0]{}%
\providecommand \url  [0]{\begingroup\@sanitize@url \@url }%
\providecommand \@url [1]{\endgroup\@href {#1}{\urlprefix }}%
\providecommand \urlprefix  [0]{URL }%
\providecommand \Eprint [0]{\href }%
\providecommand \doibase [0]{http://dx.doi.org/}%
\providecommand \selectlanguage [0]{\@gobble}%
\providecommand \bibinfo  [0]{\@secondoftwo}%
\providecommand \bibfield  [0]{\@secondoftwo}%
\providecommand \translation [1]{[#1]}%
\providecommand \BibitemOpen [0]{}%
\providecommand \bibitemStop [0]{}%
\providecommand \bibitemNoStop [0]{.\EOS\space}%
\providecommand \EOS [0]{\spacefactor3000\relax}%
\providecommand \BibitemShut  [1]{\csname bibitem#1\endcsname}%
\let\auto@bib@innerbib\@empty
\bibitem [{\citenamefont {{Fukuda}}\ \emph {et~al.}(1998)\citenamefont
  {{Fukuda}} \emph {et~al.}}]{SuperK_1998}%
  \BibitemOpen
  \bibfield  {author} {\bibinfo {author} {\bibfnamefont {Y.}~\bibnamefont
  {{Fukuda}}} \emph {et~al.},\ }\href {\doibase 10.1103/PhysRevLett.81.1562}
  {\bibfield  {journal} {\bibinfo  {journal} {\prl}\ }\textbf {\bibinfo
  {volume} {81}},\ \bibinfo {pages} {1562} (\bibinfo {year} {1998})},\ \Eprint
  {http://arxiv.org/abs/hep-ex/9807003} {arXiv:hep-ex/9807003 [hep-ex]}
  \BibitemShut {NoStop}%
\bibitem [{\citenamefont {{SNO Collaboration}}(2001)}]{SNO_2001}%
  \BibitemOpen
  \bibfield  {author} {\bibinfo {author} {\bibnamefont {{SNO Collaboration}}},\
  }\href {\doibase 10.1103/PhysRevLett.87.071301} {\bibfield  {journal}
  {\bibinfo  {journal} {Phys. Rev. Lett.}\ }\textbf {\bibinfo {volume} {87}},\
  \bibinfo {pages} {071301} (\bibinfo {year} {2001})}\BibitemShut {NoStop}%
\bibitem [{\citenamefont {{SNO Collaboration}}(2002)}]{SNO_2002}%
  \BibitemOpen
  \bibfield  {author} {\bibinfo {author} {\bibnamefont {{SNO Collaboration}}},\
  }\href {\doibase 10.1103/PhysRevLett.89.011301} {\bibfield  {journal}
  {\bibinfo  {journal} {Phys. Rev. Lett.}\ }\textbf {\bibinfo {volume} {89}},\
  \bibinfo {pages} {011301} (\bibinfo {year} {2002})}\BibitemShut {NoStop}%
\bibitem [{\citenamefont {{Esteban}}\ \emph {et~al.}(2019)\citenamefont
  {{Esteban}}, \citenamefont {{Gonzalez-Garcia}}, \citenamefont
  {{Hernandez-Cabezudo}}, \citenamefont {{Maltoni}},\ and\ \citenamefont
  {{Schwetz}}}]{2019JHEP...01..106E}%
  \BibitemOpen
  \bibfield  {author} {\bibinfo {author} {\bibfnamefont {I.}~\bibnamefont
  {{Esteban}}}, \bibinfo {author} {\bibfnamefont {M.~C.}\ \bibnamefont
  {{Gonzalez-Garcia}}}, \bibinfo {author} {\bibfnamefont {A.}~\bibnamefont
  {{Hernandez-Cabezudo}}}, \bibinfo {author} {\bibfnamefont {M.}~\bibnamefont
  {{Maltoni}}}, \ and\ \bibinfo {author} {\bibfnamefont {T.}~\bibnamefont
  {{Schwetz}}},\ }\href {\doibase 10.1007/JHEP01(2019)106} {\bibfield
  {journal} {\bibinfo  {journal} {Journal of High Energy Physics}\ }\textbf
  {\bibinfo {volume} {2019}},\ \bibinfo {eid} {106} (\bibinfo {year} {2019})},\
  \Eprint {http://arxiv.org/abs/1811.05487} {arXiv:1811.05487 [hep-ph]}
  \BibitemShut {NoStop}%
\bibitem [{\citenamefont {Aker}\ \emph {et~al.}(2022)\citenamefont {Aker} \emph
  {et~al.}}]{aker2021direct}%
  \BibitemOpen
  \bibfield  {author} {\bibinfo {author} {\bibfnamefont {M.}~\bibnamefont
  {Aker}} \emph {et~al.} (\bibinfo {collaboration} {KATRIN}),\ }\href {\doibase
  10.1038/s41567-021-01463-1} {\bibfield  {journal} {\bibinfo  {journal}
  {Nature Phys.}\ }\textbf {\bibinfo {volume} {18}},\ \bibinfo {pages} {160}
  (\bibinfo {year} {2022})},\ \Eprint {http://arxiv.org/abs/2105.08533}
  {arXiv:2105.08533 [hep-ex]} \BibitemShut {NoStop}%
\bibitem [{\citenamefont {{KATRIN Collaboration}}(2005)}]{KATRIN_report}%
  \BibitemOpen
  \bibfield  {author} {\bibinfo {author} {\bibnamefont {{KATRIN
  Collaboration}}},\ }\href {\doibase 10.5445/IR/270060419} {\emph {\bibinfo
  {title} {KATRIN design report 2004}}},\ \bibinfo {type} {Tech. Rep.}\
  (\bibinfo  {institution} {{Forschungszentrum Jülich}},\ \bibinfo {year}
  {2005})\ \bibinfo {note} {51.54.01; LK 01}\BibitemShut {NoStop}%
\bibitem [{\citenamefont {Lesgourgues}\ \emph {et~al.}(2004)\citenamefont
  {Lesgourgues}, \citenamefont {Pastor},\ and\ \citenamefont
  {Perotto}}]{PhysRevD.70.045016}%
  \BibitemOpen
  \bibfield  {author} {\bibinfo {author} {\bibfnamefont {J.}~\bibnamefont
  {Lesgourgues}}, \bibinfo {author} {\bibfnamefont {S.}~\bibnamefont {Pastor}},
  \ and\ \bibinfo {author} {\bibfnamefont {L.}~\bibnamefont {Perotto}},\ }\href
  {\doibase 10.1103/PhysRevD.70.045016} {\bibfield  {journal} {\bibinfo
  {journal} {Phys. Rev. D}\ }\textbf {\bibinfo {volume} {70}},\ \bibinfo
  {pages} {045016} (\bibinfo {year} {2004})}\BibitemShut {NoStop}%
\bibitem [{\citenamefont {Lesgourgues}\ \emph {et~al.}(2006)\citenamefont
  {Lesgourgues}, \citenamefont {Perotto}, \citenamefont {Pastor},\ and\
  \citenamefont {Piat}}]{Lesgourgues_2006}%
  \BibitemOpen
  \bibfield  {author} {\bibinfo {author} {\bibfnamefont {J.}~\bibnamefont
  {Lesgourgues}}, \bibinfo {author} {\bibfnamefont {L.}~\bibnamefont
  {Perotto}}, \bibinfo {author} {\bibfnamefont {S.}~\bibnamefont {Pastor}}, \
  and\ \bibinfo {author} {\bibfnamefont {M.}~\bibnamefont {Piat}},\ }\href
  {\doibase 10.1103/physrevd.73.045021} {\bibfield  {journal} {\bibinfo
  {journal} {Physical Review D}\ }\textbf {\bibinfo {volume} {73}} (\bibinfo
  {year} {2006}),\ 10.1103/physrevd.73.045021}\BibitemShut {NoStop}%
\bibitem [{\citenamefont {Jimenez}\ \emph {et~al.}(2010)\citenamefont
  {Jimenez}, \citenamefont {Kitching}, \citenamefont {Peña-Garay},\ and\
  \citenamefont {Verde}}]{Jimenez_2010}%
  \BibitemOpen
  \bibfield  {author} {\bibinfo {author} {\bibfnamefont {R.}~\bibnamefont
  {Jimenez}}, \bibinfo {author} {\bibfnamefont {T.}~\bibnamefont {Kitching}},
  \bibinfo {author} {\bibfnamefont {C.}~\bibnamefont {Peña-Garay}}, \ and\
  \bibinfo {author} {\bibfnamefont {L.}~\bibnamefont {Verde}},\ }\href
  {\doibase 10.1088/1475-7516/2010/05/035} {\bibfield  {journal} {\bibinfo
  {journal} {Journal of Cosmology and Astroparticle Physics}\ }\textbf
  {\bibinfo {volume} {2010}},\ \bibinfo {pages} {035–035} (\bibinfo {year}
  {2010})}\BibitemShut {NoStop}%
\bibitem [{\citenamefont {Hamann}\ \emph {et~al.}(2012)\citenamefont {Hamann},
  \citenamefont {Hannestad},\ and\ \citenamefont {Wong}}]{Hamann_2012}%
  \BibitemOpen
  \bibfield  {author} {\bibinfo {author} {\bibfnamefont {J.}~\bibnamefont
  {Hamann}}, \bibinfo {author} {\bibfnamefont {S.}~\bibnamefont {Hannestad}}, \
  and\ \bibinfo {author} {\bibfnamefont {Y.~Y.}\ \bibnamefont {Wong}},\ }\href
  {\doibase 10.1088/1475-7516/2012/11/052} {\bibfield  {journal} {\bibinfo
  {journal} {Journal of Cosmology and Astroparticle Physics}\ }\textbf
  {\bibinfo {volume} {2012}},\ \bibinfo {pages} {052} (\bibinfo {year}
  {2012})}\BibitemShut {NoStop}%
\bibitem [{\citenamefont {Hannestad}\ and\ \citenamefont
  {Schwetz}(2016)}]{Hannestad_2016}%
  \BibitemOpen
  \bibfield  {author} {\bibinfo {author} {\bibfnamefont {S.}~\bibnamefont
  {Hannestad}}\ and\ \bibinfo {author} {\bibfnamefont {T.}~\bibnamefont
  {Schwetz}},\ }\href {\doibase 10.1088/1475-7516/2016/11/035} {\bibfield
  {journal} {\bibinfo  {journal} {Journal of Cosmology and Astroparticle
  Physics}\ }\textbf {\bibinfo {volume} {2016}},\ \bibinfo {pages} {035–035}
  (\bibinfo {year} {2016})}\BibitemShut {NoStop}%
\bibitem [{\citenamefont {Vagnozzi}\ \emph {et~al.}(2018)\citenamefont
  {Vagnozzi}, \citenamefont {Dhawan}, \citenamefont {Gerbino}, \citenamefont
  {Freese}, \citenamefont {Goobar},\ and\ \citenamefont
  {Mena}}]{Vagnozzi_2018}%
  \BibitemOpen
  \bibfield  {author} {\bibinfo {author} {\bibfnamefont {S.}~\bibnamefont
  {Vagnozzi}}, \bibinfo {author} {\bibfnamefont {S.}~\bibnamefont {Dhawan}},
  \bibinfo {author} {\bibfnamefont {M.}~\bibnamefont {Gerbino}}, \bibinfo
  {author} {\bibfnamefont {K.}~\bibnamefont {Freese}}, \bibinfo {author}
  {\bibfnamefont {A.}~\bibnamefont {Goobar}}, \ and\ \bibinfo {author}
  {\bibfnamefont {O.}~\bibnamefont {Mena}},\ }\href {\doibase
  10.1103/physrevd.98.083501} {\bibfield  {journal} {\bibinfo  {journal}
  {Physical Review D}\ }\textbf {\bibinfo {volume} {98}} (\bibinfo {year}
  {2018}),\ 10.1103/physrevd.98.083501}\BibitemShut {NoStop}%
\bibitem [{\citenamefont {Collaboration}(2020)}]{planck_pr3}%
  \BibitemOpen
  \bibfield  {author} {\bibinfo {author} {\bibfnamefont {P.}~\bibnamefont
  {Collaboration}},\ }\href {\doibase 10.1051/0004-6361/201936386} {\bibfield
  {journal} {\bibinfo  {journal} {Astronomy Astrophysics}\ }\textbf {\bibinfo
  {volume} {641}},\ \bibinfo {pages} {A5} (\bibinfo {year} {2020})}\BibitemShut
  {NoStop}%
\bibitem [{\citenamefont {{Choudhury}}\ and\ \citenamefont
  {{Hannestad}}(2020)}]{Choudhury_2020}%
  \BibitemOpen
  \bibfield  {author} {\bibinfo {author} {\bibfnamefont {S.~R.}\ \bibnamefont
  {{Choudhury}}}\ and\ \bibinfo {author} {\bibfnamefont {S.}~\bibnamefont
  {{Hannestad}}},\ }\href {\doibase 10.1088/1475-7516/2020/07/037} {\bibfield
  {journal} {\bibinfo  {journal} {\jcap}\ }\textbf {\bibinfo {volume} {2020}},\
  \bibinfo {eid} {037} (\bibinfo {year} {2020})},\ \Eprint
  {http://arxiv.org/abs/1907.12598} {arXiv:1907.12598 [astro-ph.CO]}
  \BibitemShut {NoStop}%
\bibitem [{\citenamefont {Wang}\ \emph {et~al.}(2024)\citenamefont {Wang},
  \citenamefont {Mena}, \citenamefont {Valentino},\ and\ \citenamefont
  {Gariazzo}}]{wang2024}%
  \BibitemOpen
  \bibfield  {author} {\bibinfo {author} {\bibfnamefont {D.}~\bibnamefont
  {Wang}}, \bibinfo {author} {\bibfnamefont {O.}~\bibnamefont {Mena}}, \bibinfo
  {author} {\bibfnamefont {E.~D.}\ \bibnamefont {Valentino}}, \ and\ \bibinfo
  {author} {\bibfnamefont {S.}~\bibnamefont {Gariazzo}},\ }\href
  {https://arxiv.org/abs/2405.03368} {\enquote {\bibinfo {title} {Updating
  neutrino mass constraints with background measurements},}\ } (\bibinfo {year}
  {2024}),\ \Eprint {http://arxiv.org/abs/2405.03368} {arXiv:2405.03368
  [astro-ph.CO]} \BibitemShut {NoStop}%
\bibitem [{\citenamefont {Anderson}\ \emph {et~al.}(2012)\citenamefont
  {Anderson} \emph {et~al.}}]{Anderson_2012}%
  \BibitemOpen
  \bibfield  {author} {\bibinfo {author} {\bibfnamefont {L.}~\bibnamefont
  {Anderson}} \emph {et~al.},\ }\href {\doibase
  10.1111/j.1365-2966.2012.22066.x} {\bibfield  {journal} {\bibinfo  {journal}
  {Monthly Notices of the Royal Astronomical Society}\ }\textbf {\bibinfo
  {volume} {427}},\ \bibinfo {pages} {3435} (\bibinfo {year}
  {2012})}\BibitemShut {NoStop}%
\bibitem [{\citenamefont {{Planck Collaboration}}(2020)}]{planck18}%
  \BibitemOpen
  \bibfield  {author} {\bibinfo {author} {\bibnamefont {{Planck
  Collaboration}}},\ }\href {\doibase 10.1051/0004-6361/201833910} {\bibfield
  {journal} {\bibinfo  {journal} {Astronomy Astrophysics}\ }\textbf {\bibinfo
  {volume} {641}},\ \bibinfo {pages} {A6} (\bibinfo {year} {2020})}\BibitemShut
  {NoStop}%
\bibitem [{\citenamefont {{Madhavacheril}}\ \emph {et~al.}(2024)\citenamefont
  {{Madhavacheril}}, \citenamefont {{Qu}}, \citenamefont {{Sherwin}},
  \citenamefont {{MacCrann}}, \citenamefont {{Li}} \emph {et~al.}}]{act_dr6}%
  \BibitemOpen
  \bibfield  {author} {\bibinfo {author} {\bibfnamefont {M.~S.}\ \bibnamefont
  {{Madhavacheril}}}, \bibinfo {author} {\bibfnamefont {F.~J.}\ \bibnamefont
  {{Qu}}}, \bibinfo {author} {\bibfnamefont {B.~D.}\ \bibnamefont {{Sherwin}}},
  \bibinfo {author} {\bibfnamefont {N.}~\bibnamefont {{MacCrann}}}, \bibinfo
  {author} {\bibfnamefont {Y.}~\bibnamefont {{Li}}},  \emph {et~al.},\ }\href
  {\doibase 10.3847/1538-4357/acff5f} {\bibfield  {journal} {\bibinfo
  {journal} {\apj}\ }\textbf {\bibinfo {volume} {962}},\ \bibinfo {eid} {113}
  (\bibinfo {year} {2024})},\ \Eprint {http://arxiv.org/abs/2304.05203}
  {arXiv:2304.05203 [astro-ph.CO]} \BibitemShut {NoStop}%
\bibitem [{\citenamefont {Allali}\ and\ \citenamefont
  {Notari}(2024)}]{pr4_desi_sdss}%
  \BibitemOpen
  \bibfield  {author} {\bibinfo {author} {\bibfnamefont {I.~J.}\ \bibnamefont
  {Allali}}\ and\ \bibinfo {author} {\bibfnamefont {A.}~\bibnamefont
  {Notari}},\ }\href {https://arxiv.org/abs/2406.14554} {} (\bibinfo {year}
  {2024}),\ \Eprint {http://arxiv.org/abs/2406.14554} {arXiv:2406.14554
  [astro-ph.CO]} \BibitemShut {NoStop}%
\bibitem [{\citenamefont {{DESI Collaboration}}(2024{\natexlab{a}})}]{desi}%
  \BibitemOpen
  \bibfield  {author} {\bibinfo {author} {\bibnamefont {{DESI
  Collaboration}}},\ }\href@noop {} {} (\bibinfo {year} {2024}{\natexlab{a}}),\
  \Eprint {http://arxiv.org/abs/2404.03002} {arXiv:2404.03002 [astro-ph.CO]}
  \BibitemShut {NoStop}%
\bibitem [{\citenamefont {Allison}\ \emph {et~al.}(2015)\citenamefont
  {Allison}, \citenamefont {Caucal}, \citenamefont {Calabrese}, \citenamefont
  {Dunkley},\ and\ \citenamefont {Louis}}]{Allison_2015}%
  \BibitemOpen
  \bibfield  {author} {\bibinfo {author} {\bibfnamefont {R.}~\bibnamefont
  {Allison}}, \bibinfo {author} {\bibfnamefont {P.}~\bibnamefont {Caucal}},
  \bibinfo {author} {\bibfnamefont {E.}~\bibnamefont {Calabrese}}, \bibinfo
  {author} {\bibfnamefont {J.}~\bibnamefont {Dunkley}}, \ and\ \bibinfo
  {author} {\bibfnamefont {T.}~\bibnamefont {Louis}},\ }\href {\doibase
  10.1103/physrevd.92.123535} {\bibfield  {journal} {\bibinfo  {journal}
  {Physical Review D}\ }\textbf {\bibinfo {volume} {92}} (\bibinfo {year}
  {2015}),\ 10.1103/physrevd.92.123535}\BibitemShut {NoStop}%
\bibitem [{\citenamefont {Choudhury}\ and\ \citenamefont
  {Choubey}(2018)}]{Choudhury_2018}%
  \BibitemOpen
  \bibfield  {author} {\bibinfo {author} {\bibfnamefont {S.~R.}\ \bibnamefont
  {Choudhury}}\ and\ \bibinfo {author} {\bibfnamefont {S.}~\bibnamefont
  {Choubey}},\ }\href {\doibase 10.1088/1475-7516/2018/09/017} {\bibfield
  {journal} {\bibinfo  {journal} {Journal of Cosmology and Astroparticle
  Physics}\ }\textbf {\bibinfo {volume} {2018}},\ \bibinfo {pages} {017}
  (\bibinfo {year} {2018})}\BibitemShut {NoStop}%
\bibitem [{\citenamefont {Elbers}\ \emph {et~al.}(2024)\citenamefont {Elbers},
  \citenamefont {Frenk}, \citenamefont {Jenkins}, \citenamefont {Li},\ and\
  \citenamefont {Pascoli}}]{elbers2024}%
  \BibitemOpen
  \bibfield  {author} {\bibinfo {author} {\bibfnamefont {W.}~\bibnamefont
  {Elbers}}, \bibinfo {author} {\bibfnamefont {C.~S.}\ \bibnamefont {Frenk}},
  \bibinfo {author} {\bibfnamefont {A.}~\bibnamefont {Jenkins}}, \bibinfo
  {author} {\bibfnamefont {B.}~\bibnamefont {Li}}, \ and\ \bibinfo {author}
  {\bibfnamefont {S.}~\bibnamefont {Pascoli}},\ }\href
  {https://arxiv.org/abs/2407.10965} {\enquote {\bibinfo {title} {Negative
  neutrino masses as a mirage of dark energy},}\ } (\bibinfo {year} {2024}),\
  \Eprint {http://arxiv.org/abs/2407.10965} {arXiv:2407.10965 [astro-ph.CO]}
  \BibitemShut {NoStop}%
\bibitem [{\citenamefont {Naredo-Tuero}\ \emph {et~al.}(2024)\citenamefont
  {Naredo-Tuero}, \citenamefont {Escudero}, \citenamefont
  {Fernández-Martínez}, \citenamefont {Marcano},\ and\ \citenamefont
  {Poulin}}]{naredotuero2024}%
  \BibitemOpen
  \bibfield  {author} {\bibinfo {author} {\bibfnamefont {D.}~\bibnamefont
  {Naredo-Tuero}}, \bibinfo {author} {\bibfnamefont {M.}~\bibnamefont
  {Escudero}}, \bibinfo {author} {\bibfnamefont {E.}~\bibnamefont
  {Fernández-Martínez}}, \bibinfo {author} {\bibfnamefont {X.}~\bibnamefont
  {Marcano}}, \ and\ \bibinfo {author} {\bibfnamefont {V.}~\bibnamefont
  {Poulin}},\ }\href {https://arxiv.org/abs/2407.13831} {\enquote {\bibinfo
  {title} {Living at the edge: A critical look at the cosmological neutrino
  mass bound},}\ } (\bibinfo {year} {2024}),\ \Eprint
  {http://arxiv.org/abs/2407.13831} {arXiv:2407.13831 [astro-ph.CO]}
  \BibitemShut {NoStop}%
\bibitem [{\citenamefont {{Qu}}\ \emph {et~al.}(2024)\citenamefont {{Qu}},
  \citenamefont {{Sherwin}}, \citenamefont {{Madhavacheril}}, \citenamefont
  {{Han}}, \citenamefont {{Crowley}} \emph {et~al.}}]{qu2023atacama}%
  \BibitemOpen
  \bibfield  {author} {\bibinfo {author} {\bibfnamefont {F.~J.}\ \bibnamefont
  {{Qu}}}, \bibinfo {author} {\bibfnamefont {B.~D.}\ \bibnamefont {{Sherwin}}},
  \bibinfo {author} {\bibfnamefont {M.~S.}\ \bibnamefont {{Madhavacheril}}},
  \bibinfo {author} {\bibfnamefont {D.}~\bibnamefont {{Han}}}, \bibinfo
  {author} {\bibfnamefont {K.~T.}\ \bibnamefont {{Crowley}}},  \emph {et~al.},\
  }\href {\doibase 10.3847/1538-4357/acfe06} {\bibfield  {journal} {\bibinfo
  {journal} {\apj}\ }\textbf {\bibinfo {volume} {962}},\ \bibinfo {eid} {112}
  (\bibinfo {year} {2024})},\ \Eprint {http://arxiv.org/abs/2304.05202}
  {arXiv:2304.05202 [astro-ph.CO]} \BibitemShut {NoStop}%
\bibitem [{\citenamefont {Brout}\ \emph {et~al.}(2022)\citenamefont {Brout},
  \citenamefont {Scolnic}, \citenamefont {Popovic}, \citenamefont {Riess},
  \citenamefont {Carr}, \citenamefont {Zuntz}, \citenamefont {Kessler},
  \citenamefont {Davis}, \citenamefont {Hinton}, \citenamefont {Jones},
  \citenamefont {Kenworthy}, \citenamefont {Peterson}, \citenamefont {Said}
  \emph {et~al.}}]{pantheonp}%
  \BibitemOpen
  \bibfield  {author} {\bibinfo {author} {\bibfnamefont {D.}~\bibnamefont
  {Brout}}, \bibinfo {author} {\bibfnamefont {D.}~\bibnamefont {Scolnic}},
  \bibinfo {author} {\bibfnamefont {B.}~\bibnamefont {Popovic}}, \bibinfo
  {author} {\bibfnamefont {A.~G.}\ \bibnamefont {Riess}}, \bibinfo {author}
  {\bibfnamefont {A.}~\bibnamefont {Carr}}, \bibinfo {author} {\bibfnamefont
  {J.}~\bibnamefont {Zuntz}}, \bibinfo {author} {\bibfnamefont
  {R.}~\bibnamefont {Kessler}}, \bibinfo {author} {\bibfnamefont {T.~M.}\
  \bibnamefont {Davis}}, \bibinfo {author} {\bibfnamefont {S.}~\bibnamefont
  {Hinton}}, \bibinfo {author} {\bibfnamefont {D.}~\bibnamefont {Jones}},
  \bibinfo {author} {\bibfnamefont {W.~D.}\ \bibnamefont {Kenworthy}}, \bibinfo
  {author} {\bibfnamefont {E.~R.}\ \bibnamefont {Peterson}}, \bibinfo {author}
  {\bibfnamefont {K.}~\bibnamefont {Said}},  \emph {et~al.},\ }\href {\doibase
  10.3847/1538-4357/ac8e04} {\bibfield  {journal} {\bibinfo  {journal} {The
  Astrophysical Journal}\ }\textbf {\bibinfo {volume} {938}},\ \bibinfo {pages}
  {110} (\bibinfo {year} {2022})}\BibitemShut {NoStop}%
\bibitem [{\citenamefont {{Rosenberg}}\ \emph {et~al.}(2022)\citenamefont
  {{Rosenberg}}, \citenamefont {{Gratton}},\ and\ \citenamefont
  {{Efstathiou}}}]{rosenberg22}%
  \BibitemOpen
  \bibfield  {author} {\bibinfo {author} {\bibfnamefont {E.}~\bibnamefont
  {{Rosenberg}}}, \bibinfo {author} {\bibfnamefont {S.}~\bibnamefont
  {{Gratton}}}, \ and\ \bibinfo {author} {\bibfnamefont {G.}~\bibnamefont
  {{Efstathiou}}},\ }\href {\doibase 10.1093/mnras/stac2744} {\bibfield
  {journal} {\bibinfo  {journal} {\mnras}\ }\textbf {\bibinfo {volume} {517}},\
  \bibinfo {pages} {4620} (\bibinfo {year} {2022})},\ \Eprint
  {http://arxiv.org/abs/2205.10869} {arXiv:2205.10869 [astro-ph.CO]}
  \BibitemShut {NoStop}%
\bibitem [{\citenamefont {Beutler}\ \emph {et~al.}(2016)\citenamefont
  {Beutler}, \citenamefont {Seo}, \citenamefont {Saito}, \citenamefont
  {Chuang}, \citenamefont {Cuesta}, \citenamefont {Eisenstein}, \citenamefont
  {Gil-Mar{\'{\i}}n}, \citenamefont {Grieb}, \citenamefont {Hand},
  \citenamefont {Kitaura}, \citenamefont {Modi}, \citenamefont {Nichol},
  \citenamefont {Olmstead}, \citenamefont {Percival}, \citenamefont {Prada},
  \citenamefont {S{\'{a}}nchez}, \citenamefont {Rodriguez-Torres},
  \citenamefont {Ross}, \citenamefont {Ross}, \citenamefont {Schneider},
  \citenamefont {Tinker}, \citenamefont {Tojeiro},\ and\ \citenamefont
  {Vargas-Maga{\~{n}}a}}]{Beutler_2016}%
  \BibitemOpen
  \bibfield  {author} {\bibinfo {author} {\bibfnamefont {F.}~\bibnamefont
  {Beutler}}, \bibinfo {author} {\bibfnamefont {H.-J.}\ \bibnamefont {Seo}},
  \bibinfo {author} {\bibfnamefont {S.}~\bibnamefont {Saito}}, \bibinfo
  {author} {\bibfnamefont {C.-H.}\ \bibnamefont {Chuang}}, \bibinfo {author}
  {\bibfnamefont {A.~J.}\ \bibnamefont {Cuesta}}, \bibinfo {author}
  {\bibfnamefont {D.~J.}\ \bibnamefont {Eisenstein}}, \bibinfo {author}
  {\bibfnamefont {H.}~\bibnamefont {Gil-Mar{\'{\i}}n}}, \bibinfo {author}
  {\bibfnamefont {J.~N.}\ \bibnamefont {Grieb}}, \bibinfo {author}
  {\bibfnamefont {N.}~\bibnamefont {Hand}}, \bibinfo {author} {\bibfnamefont
  {F.-S.}\ \bibnamefont {Kitaura}}, \bibinfo {author} {\bibfnamefont
  {C.}~\bibnamefont {Modi}}, \bibinfo {author} {\bibfnamefont {R.~C.}\
  \bibnamefont {Nichol}}, \bibinfo {author} {\bibfnamefont {M.~D.}\
  \bibnamefont {Olmstead}}, \bibinfo {author} {\bibfnamefont {W.~J.}\
  \bibnamefont {Percival}}, \bibinfo {author} {\bibfnamefont {F.}~\bibnamefont
  {Prada}}, \bibinfo {author} {\bibfnamefont {A.~G.}\ \bibnamefont
  {S{\'{a}}nchez}}, \bibinfo {author} {\bibfnamefont {S.}~\bibnamefont
  {Rodriguez-Torres}}, \bibinfo {author} {\bibfnamefont {A.~J.}\ \bibnamefont
  {Ross}}, \bibinfo {author} {\bibfnamefont {N.~P.}\ \bibnamefont {Ross}},
  \bibinfo {author} {\bibfnamefont {D.~P.}\ \bibnamefont {Schneider}}, \bibinfo
  {author} {\bibfnamefont {J.}~\bibnamefont {Tinker}}, \bibinfo {author}
  {\bibfnamefont {R.}~\bibnamefont {Tojeiro}}, \ and\ \bibinfo {author}
  {\bibfnamefont {M.}~\bibnamefont {Vargas-Maga{\~{n}}a}},\ }\href {\doibase
  10.1093/mnras/stw3298} {\bibfield  {journal} {\bibinfo  {journal} {Monthly
  Notices of the Royal Astronomical Society}\ }\textbf {\bibinfo {volume}
  {466}},\ \bibinfo {pages} {2242} (\bibinfo {year} {2016})}\BibitemShut
  {NoStop}%
\bibitem [{\citenamefont {Percival}\ \emph {et~al.}(2010)\citenamefont
  {Percival}, \citenamefont {Reid}, \citenamefont {Eisenstein} \emph
  {et~al.}}]{Percival_2010}%
  \BibitemOpen
  \bibfield  {author} {\bibinfo {author} {\bibfnamefont {W.~J.}\ \bibnamefont
  {Percival}}, \bibinfo {author} {\bibfnamefont {B.~A.}\ \bibnamefont {Reid}},
  \bibinfo {author} {\bibfnamefont {D.~J.}\ \bibnamefont {Eisenstein}},  \emph
  {et~al.},\ }\href {\doibase 10.1111/j.1365-2966.2009.15812.x} {\bibfield
  {journal} {\bibinfo  {journal} {Monthly Notices of the Royal Astronomical
  Society}\ }\textbf {\bibinfo {volume} {401}},\ \bibinfo {pages} {2148}
  (\bibinfo {year} {2010})}\BibitemShut {NoStop}%
\bibitem [{\citenamefont {{Dawson}}\ \emph {et~al.}(2013)\citenamefont
  {{Dawson}}, \citenamefont {{Schlegel}} \emph {et~al.}}]{BOSS_bao}%
  \BibitemOpen
  \bibfield  {author} {\bibinfo {author} {\bibfnamefont {K.~S.}\ \bibnamefont
  {{Dawson}}}, \bibinfo {author} {\bibfnamefont {D.~J.}\ \bibnamefont
  {{Schlegel}}},  \emph {et~al.},\ }\href {\doibase 10.1088/0004-6256/145/1/10}
  {\bibfield  {journal} {\bibinfo  {journal} {\aj}\ }\textbf {\bibinfo {volume}
  {145}},\ \bibinfo {eid} {10} (\bibinfo {year} {2013})},\ \Eprint
  {http://arxiv.org/abs/1208.0022} {arXiv:1208.0022 [astro-ph.CO]} \BibitemShut
  {NoStop}%
\bibitem [{\citenamefont {Torrado}\ and\ \citenamefont
  {Lewis}(2021)}]{Torrado_2021}%
  \BibitemOpen
  \bibfield  {author} {\bibinfo {author} {\bibfnamefont {J.}~\bibnamefont
  {Torrado}}\ and\ \bibinfo {author} {\bibfnamefont {A.}~\bibnamefont
  {Lewis}},\ }\href {\doibase 10.1088/1475-7516/2021/05/057} {\bibfield
  {journal} {\bibinfo  {journal} {Journal of Cosmology and Astroparticle
  Physics}\ }\textbf {\bibinfo {volume} {2021}},\ \bibinfo {pages} {057}
  (\bibinfo {year} {2021})}\BibitemShut {NoStop}%
\bibitem [{\citenamefont {Carron}\ \emph {et~al.}(2022)\citenamefont {Carron},
  \citenamefont {Mirmelstein},\ and\ \citenamefont {Lewis}}]{pr4_lens}%
  \BibitemOpen
  \bibfield  {author} {\bibinfo {author} {\bibfnamefont {J.}~\bibnamefont
  {Carron}}, \bibinfo {author} {\bibfnamefont {M.}~\bibnamefont {Mirmelstein}},
  \ and\ \bibinfo {author} {\bibfnamefont {A.}~\bibnamefont {Lewis}},\ }\href
  {\doibase 10.1088/1475-7516/2022/09/039} {\bibfield  {journal} {\bibinfo
  {journal} {Journal of Cosmology and Astroparticle Physics}\ }\textbf
  {\bibinfo {volume} {2022}},\ \bibinfo {pages} {039} (\bibinfo {year}
  {2022})}\BibitemShut {NoStop}%
\bibitem [{\citenamefont {Riess}\ \emph {et~al.}(2022)\citenamefont {Riess},
  \citenamefont {Yuan}, \citenamefont {Macri}, \citenamefont {Scolnic},
  \citenamefont {Brout}, \citenamefont {Casertano}, \citenamefont {Jones},
  \citenamefont {Murakami}, \citenamefont {Anand}, \citenamefont {Breuval},
  \citenamefont {Brink}, \citenamefont {Filippenko}, \citenamefont {Hoffmann},
  \citenamefont {Jha}, \citenamefont {D’arcy~Kenworthy}, \citenamefont
  {Mackenty}, \citenamefont {Stahl},\ and\ \citenamefont {Zheng}}]{Riess_2022}%
  \BibitemOpen
  \bibfield  {author} {\bibinfo {author} {\bibfnamefont {A.~G.}\ \bibnamefont
  {Riess}}, \bibinfo {author} {\bibfnamefont {W.}~\bibnamefont {Yuan}},
  \bibinfo {author} {\bibfnamefont {L.~M.}\ \bibnamefont {Macri}}, \bibinfo
  {author} {\bibfnamefont {D.}~\bibnamefont {Scolnic}}, \bibinfo {author}
  {\bibfnamefont {D.}~\bibnamefont {Brout}}, \bibinfo {author} {\bibfnamefont
  {S.}~\bibnamefont {Casertano}}, \bibinfo {author} {\bibfnamefont {D.~O.}\
  \bibnamefont {Jones}}, \bibinfo {author} {\bibfnamefont {Y.}~\bibnamefont
  {Murakami}}, \bibinfo {author} {\bibfnamefont {G.~S.}\ \bibnamefont {Anand}},
  \bibinfo {author} {\bibfnamefont {L.}~\bibnamefont {Breuval}}, \bibinfo
  {author} {\bibfnamefont {T.~G.}\ \bibnamefont {Brink}}, \bibinfo {author}
  {\bibfnamefont {A.~V.}\ \bibnamefont {Filippenko}}, \bibinfo {author}
  {\bibfnamefont {S.}~\bibnamefont {Hoffmann}}, \bibinfo {author}
  {\bibfnamefont {S.~W.}\ \bibnamefont {Jha}}, \bibinfo {author} {\bibfnamefont
  {W.}~\bibnamefont {D’arcy~Kenworthy}}, \bibinfo {author} {\bibfnamefont
  {J.}~\bibnamefont {Mackenty}}, \bibinfo {author} {\bibfnamefont {B.~E.}\
  \bibnamefont {Stahl}}, \ and\ \bibinfo {author} {\bibfnamefont
  {W.}~\bibnamefont {Zheng}},\ }\href {\doibase 10.3847/2041-8213/ac5c5b}
  {\bibfield  {journal} {\bibinfo  {journal} {The Astrophysical Journal
  Letters}\ }\textbf {\bibinfo {volume} {934}},\ \bibinfo {pages} {L7}
  (\bibinfo {year} {2022})}\BibitemShut {NoStop}%
\bibitem [{\citenamefont {Bidenko}\ \emph {et~al.}(2023)\citenamefont
  {Bidenko}, \citenamefont {Koopmans},\ and\ \citenamefont
  {Meerburg}}]{bidenko}%
  \BibitemOpen
  \bibfield  {author} {\bibinfo {author} {\bibfnamefont {B.}~\bibnamefont
  {Bidenko}}, \bibinfo {author} {\bibfnamefont {L.~V.~E.}\ \bibnamefont
  {Koopmans}}, \ and\ \bibinfo {author} {\bibfnamefont {P.~D.}\ \bibnamefont
  {Meerburg}},\ }\href@noop {} {} (\bibinfo {year} {2023}),\ \Eprint
  {http://arxiv.org/abs/2308.05157} {arXiv:2308.05157 [astro-ph.CO]}
  \BibitemShut {NoStop}%
\bibitem [{\citenamefont {Upadhye}(2019)}]{Upadhye_2019}%
  \BibitemOpen
  \bibfield  {author} {\bibinfo {author} {\bibfnamefont {A.}~\bibnamefont
  {Upadhye}},\ }\href {\doibase 10.1088/1475-7516/2019/05/041} {\bibfield
  {journal} {\bibinfo  {journal} {Journal of Cosmology and Astroparticle
  Physics}\ }\textbf {\bibinfo {volume} {2019}},\ \bibinfo {pages} {041–041}
  (\bibinfo {year} {2019})}\BibitemShut {NoStop}%
\bibitem [{\citenamefont {Hu}\ and\ \citenamefont {Sawicki}(2007)}]{ppf}%
  \BibitemOpen
  \bibfield  {author} {\bibinfo {author} {\bibfnamefont {W.}~\bibnamefont
  {Hu}}\ and\ \bibinfo {author} {\bibfnamefont {I.}~\bibnamefont {Sawicki}},\
  }\href {\doibase 10.1103/physrevd.76.104043} {\bibfield  {journal} {\bibinfo
  {journal} {Physical Review D}\ }\textbf {\bibinfo {volume} {76}} (\bibinfo
  {year} {2007}),\ 10.1103/physrevd.76.104043}\BibitemShut {NoStop}%
\bibitem [{\citenamefont {{Chevallier}}\ and\ \citenamefont
  {{Polarski}}(2001)}]{2001IJMPD..10..213C}%
  \BibitemOpen
  \bibfield  {author} {\bibinfo {author} {\bibfnamefont {M.}~\bibnamefont
  {{Chevallier}}}\ and\ \bibinfo {author} {\bibfnamefont {D.}~\bibnamefont
  {{Polarski}}},\ }\href {\doibase 10.1142/S0218271801000822} {\bibfield
  {journal} {\bibinfo  {journal} {International Journal of Modern Physics D}\
  }\textbf {\bibinfo {volume} {10}},\ \bibinfo {pages} {213} (\bibinfo {year}
  {2001})},\ \Eprint {http://arxiv.org/abs/gr-qc/0009008} {arXiv:gr-qc/0009008
  [gr-qc]} \BibitemShut {NoStop}%
\bibitem [{\citenamefont {{Linder}}(2003)}]{2003PhRvL..90i1301L}%
  \BibitemOpen
  \bibfield  {author} {\bibinfo {author} {\bibfnamefont {E.~V.}\ \bibnamefont
  {{Linder}}},\ }\href {\doibase 10.1103/PhysRevLett.90.091301} {\bibfield
  {journal} {\bibinfo  {journal} {\prl}\ }\textbf {\bibinfo {volume} {90}},\
  \bibinfo {eid} {091301} (\bibinfo {year} {2003})},\ \Eprint
  {http://arxiv.org/abs/astro-ph/0208512} {arXiv:astro-ph/0208512 [astro-ph]}
  \BibitemShut {NoStop}%
\bibitem [{\citenamefont {Smith}\ \emph {et~al.}(2003)\citenamefont {Smith},
  \citenamefont {Peacock}, \citenamefont {Jenkins}, \citenamefont {White},
  \citenamefont {Frenk}, \citenamefont {Pearce}, \citenamefont {Thomas},
  \citenamefont {Efstathiou},\ and\ \citenamefont {Couchman}}]{smith}%
  \BibitemOpen
  \bibfield  {author} {\bibinfo {author} {\bibfnamefont {R.~E.}\ \bibnamefont
  {Smith}}, \bibinfo {author} {\bibfnamefont {J.~A.}\ \bibnamefont {Peacock}},
  \bibinfo {author} {\bibfnamefont {A.}~\bibnamefont {Jenkins}}, \bibinfo
  {author} {\bibfnamefont {S.~D.~M.}\ \bibnamefont {White}}, \bibinfo {author}
  {\bibfnamefont {C.~S.}\ \bibnamefont {Frenk}}, \bibinfo {author}
  {\bibfnamefont {F.~R.}\ \bibnamefont {Pearce}}, \bibinfo {author}
  {\bibfnamefont {P.~A.}\ \bibnamefont {Thomas}}, \bibinfo {author}
  {\bibfnamefont {G.}~\bibnamefont {Efstathiou}}, \ and\ \bibinfo {author}
  {\bibfnamefont {H.~M.~P.}\ \bibnamefont {Couchman}},\ }\href {\doibase
  10.1046/j.1365-8711.2003.06503.x} {\bibfield  {journal} {\bibinfo  {journal}
  {Monthly Notices of the Royal Astronomical Society}\ }\textbf {\bibinfo
  {volume} {341}},\ \bibinfo {pages} {1311} (\bibinfo {year}
  {2003})}\BibitemShut {NoStop}%
\bibitem [{\citenamefont {Takahashi}\ \emph {et~al.}(2012)\citenamefont
  {Takahashi}, \citenamefont {Sato}, \citenamefont {Nishimichi}, \citenamefont
  {Taruya},\ and\ \citenamefont {Oguri}}]{Takahashi_2012}%
  \BibitemOpen
  \bibfield  {author} {\bibinfo {author} {\bibfnamefont {R.}~\bibnamefont
  {Takahashi}}, \bibinfo {author} {\bibfnamefont {M.}~\bibnamefont {Sato}},
  \bibinfo {author} {\bibfnamefont {T.}~\bibnamefont {Nishimichi}}, \bibinfo
  {author} {\bibfnamefont {A.}~\bibnamefont {Taruya}}, \ and\ \bibinfo {author}
  {\bibfnamefont {M.}~\bibnamefont {Oguri}},\ }\href {\doibase
  10.1088/0004-637x/761/2/152} {\bibfield  {journal} {\bibinfo  {journal} {The
  Astrophysical Journal}\ }\textbf {\bibinfo {volume} {761}},\ \bibinfo {pages}
  {152} (\bibinfo {year} {2012})}\BibitemShut {NoStop}%
\bibitem [{\citenamefont {Mead}\ \emph {et~al.}(2015)\citenamefont {Mead},
  \citenamefont {Peacock}, \citenamefont {Heymans}, \citenamefont {Joudaki},\
  and\ \citenamefont {Heavens}}]{mead_2015}%
  \BibitemOpen
  \bibfield  {author} {\bibinfo {author} {\bibfnamefont {A.~J.}\ \bibnamefont
  {Mead}}, \bibinfo {author} {\bibfnamefont {J.~A.}\ \bibnamefont {Peacock}},
  \bibinfo {author} {\bibfnamefont {C.}~\bibnamefont {Heymans}}, \bibinfo
  {author} {\bibfnamefont {S.}~\bibnamefont {Joudaki}}, \ and\ \bibinfo
  {author} {\bibfnamefont {A.~F.}\ \bibnamefont {Heavens}},\ }\href {\doibase
  10.1093/mnras/stv2036} {\bibfield  {journal} {\bibinfo  {journal} {Monthly
  Notices of the Royal Astronomical Society}\ }\textbf {\bibinfo {volume}
  {454}},\ \bibinfo {pages} {1958} (\bibinfo {year} {2015})}\BibitemShut
  {NoStop}%
\bibitem [{\citenamefont {Mahony}\ \emph {et~al.}(2020)\citenamefont {Mahony},
  \citenamefont {Leistedt}, \citenamefont {Peiris}, \citenamefont {Braden},
  \citenamefont {Joachimi}, \citenamefont {Korn}, \citenamefont {Cremonesi},\
  and\ \citenamefont {Nichol}}]{Mahony_2020}%
  \BibitemOpen
  \bibfield  {author} {\bibinfo {author} {\bibfnamefont {C.}~\bibnamefont
  {Mahony}}, \bibinfo {author} {\bibfnamefont {B.}~\bibnamefont {Leistedt}},
  \bibinfo {author} {\bibfnamefont {H.~V.}\ \bibnamefont {Peiris}}, \bibinfo
  {author} {\bibfnamefont {J.}~\bibnamefont {Braden}}, \bibinfo {author}
  {\bibfnamefont {B.}~\bibnamefont {Joachimi}}, \bibinfo {author}
  {\bibfnamefont {A.}~\bibnamefont {Korn}}, \bibinfo {author} {\bibfnamefont
  {L.}~\bibnamefont {Cremonesi}}, \ and\ \bibinfo {author} {\bibfnamefont
  {R.}~\bibnamefont {Nichol}},\ }\href {\doibase 10.1103/physrevd.101.083513}
  {\bibfield  {journal} {\bibinfo  {journal} {Physical Review D}\ }\textbf
  {\bibinfo {volume} {101}} (\bibinfo {year} {2020}),\
  10.1103/physrevd.101.083513}\BibitemShut {NoStop}%
\bibitem [{\citenamefont {Tristram}\ \emph {et~al.}(2024)\citenamefont
  {Tristram}, \citenamefont {Banday}, \citenamefont {Douspis}, \citenamefont
  {Garrido}, \citenamefont {Górski}, \citenamefont {Henrot-Versillé},
  \citenamefont {Hergt}, \citenamefont {Ilić}, \citenamefont {Keskitalo},
  \citenamefont {Lagache}, \citenamefont {Lawrence}, \citenamefont
  {Partridge},\ and\ \citenamefont {Scott}}]{Tristram_2024}%
  \BibitemOpen
  \bibfield  {author} {\bibinfo {author} {\bibfnamefont {M.}~\bibnamefont
  {Tristram}}, \bibinfo {author} {\bibfnamefont {A.~J.}\ \bibnamefont
  {Banday}}, \bibinfo {author} {\bibfnamefont {M.}~\bibnamefont {Douspis}},
  \bibinfo {author} {\bibfnamefont {X.}~\bibnamefont {Garrido}}, \bibinfo
  {author} {\bibfnamefont {K.~M.}\ \bibnamefont {Górski}}, \bibinfo {author}
  {\bibfnamefont {S.}~\bibnamefont {Henrot-Versillé}}, \bibinfo {author}
  {\bibfnamefont {L.~T.}\ \bibnamefont {Hergt}}, \bibinfo {author}
  {\bibfnamefont {S.}~\bibnamefont {Ilić}}, \bibinfo {author} {\bibfnamefont
  {R.}~\bibnamefont {Keskitalo}}, \bibinfo {author} {\bibfnamefont
  {G.}~\bibnamefont {Lagache}}, \bibinfo {author} {\bibfnamefont {C.~R.}\
  \bibnamefont {Lawrence}}, \bibinfo {author} {\bibfnamefont {B.}~\bibnamefont
  {Partridge}}, \ and\ \bibinfo {author} {\bibfnamefont {D.}~\bibnamefont
  {Scott}},\ }\href {\doibase 10.1051/0004-6361/202348015} {\bibfield
  {journal} {\bibinfo  {journal} {Astronomy \& Astrophysics}\ }\textbf
  {\bibinfo {volume} {682}},\ \bibinfo {pages} {A37} (\bibinfo {year}
  {2024})}\BibitemShut {NoStop}%
\bibitem [{\citenamefont {Efstathiou}\ and\ \citenamefont
  {Bond}(1999)}]{Efstathiou_1999}%
  \BibitemOpen
  \bibfield  {author} {\bibinfo {author} {\bibfnamefont {G.}~\bibnamefont
  {Efstathiou}}\ and\ \bibinfo {author} {\bibfnamefont {J.~R.}\ \bibnamefont
  {Bond}},\ }\href {\doibase 10.1046/j.1365-8711.1999.02274.x} {\bibfield
  {journal} {\bibinfo  {journal} {Monthly Notices of the Royal Astronomical
  Society}\ }\textbf {\bibinfo {volume} {304}},\ \bibinfo {pages} {75–97}
  (\bibinfo {year} {1999})}\BibitemShut {NoStop}%
\bibitem [{\citenamefont {Sherwin}\ \emph {et~al.}(2011)\citenamefont
  {Sherwin}, \citenamefont {Dunkley}, \citenamefont {Das} \emph
  {et~al.}}]{Sherwin_2011}%
  \BibitemOpen
  \bibfield  {author} {\bibinfo {author} {\bibfnamefont {B.~D.}\ \bibnamefont
  {Sherwin}}, \bibinfo {author} {\bibfnamefont {J.}~\bibnamefont {Dunkley}},
  \bibinfo {author} {\bibfnamefont {S.}~\bibnamefont {Das}},  \emph {et~al.},\
  }\href {\doibase 10.1103/physrevlett.107.021302} {\bibfield  {journal}
  {\bibinfo  {journal} {Physical Review Letters}\ }\textbf {\bibinfo {volume}
  {107}} (\bibinfo {year} {2011}),\ 10.1103/physrevlett.107.021302}\BibitemShut
  {NoStop}%
\bibitem [{\citenamefont {Hannestad}(2005)}]{Hannestad_2005}%
  \BibitemOpen
  \bibfield  {author} {\bibinfo {author} {\bibfnamefont {S.}~\bibnamefont
  {Hannestad}},\ }\href {\doibase 10.1103/physrevlett.95.221301} {\bibfield
  {journal} {\bibinfo  {journal} {Physical Review Letters}\ }\textbf {\bibinfo
  {volume} {95}} (\bibinfo {year} {2005}),\
  10.1103/physrevlett.95.221301}\BibitemShut {NoStop}%
\bibitem [{\citenamefont {Linder}(2003)}]{Linder_2003}%
  \BibitemOpen
  \bibfield  {author} {\bibinfo {author} {\bibfnamefont {E.~V.}\ \bibnamefont
  {Linder}},\ }\href {\doibase 10.1103/physrevlett.90.091301} {\bibfield
  {journal} {\bibinfo  {journal} {Physical Review Letters}\ }\textbf {\bibinfo
  {volume} {90}} (\bibinfo {year} {2003}),\
  10.1103/physrevlett.90.091301}\BibitemShut {NoStop}%
\bibitem [{\citenamefont {Dvorkin}\ \emph {et~al.}(2019)\citenamefont
  {Dvorkin}, \citenamefont {Gerbino}, \citenamefont {Alonso}, \citenamefont
  {Battaglia}, \citenamefont {Bird}, \citenamefont {Rivero}, \citenamefont
  {Font-Ribera}, \citenamefont {Fuller}, \citenamefont {Lattanzi},
  \citenamefont {Loverde}, \citenamefont {Muñoz}, \citenamefont {Sherwin},
  \citenamefont {Slosar},\ and\ \citenamefont
  {Villaescusa-Navarro}}]{dvorkin2020}%
  \BibitemOpen
  \bibfield  {author} {\bibinfo {author} {\bibfnamefont {C.}~\bibnamefont
  {Dvorkin}}, \bibinfo {author} {\bibfnamefont {M.}~\bibnamefont {Gerbino}},
  \bibinfo {author} {\bibfnamefont {D.}~\bibnamefont {Alonso}}, \bibinfo
  {author} {\bibfnamefont {N.}~\bibnamefont {Battaglia}}, \bibinfo {author}
  {\bibfnamefont {S.}~\bibnamefont {Bird}}, \bibinfo {author} {\bibfnamefont
  {A.~D.}\ \bibnamefont {Rivero}}, \bibinfo {author} {\bibfnamefont
  {A.}~\bibnamefont {Font-Ribera}}, \bibinfo {author} {\bibfnamefont
  {G.}~\bibnamefont {Fuller}}, \bibinfo {author} {\bibfnamefont
  {M.}~\bibnamefont {Lattanzi}}, \bibinfo {author} {\bibfnamefont
  {M.}~\bibnamefont {Loverde}}, \bibinfo {author} {\bibfnamefont {J.~B.}\
  \bibnamefont {Muñoz}}, \bibinfo {author} {\bibfnamefont {B.}~\bibnamefont
  {Sherwin}}, \bibinfo {author} {\bibfnamefont {A.}~\bibnamefont {Slosar}}, \
  and\ \bibinfo {author} {\bibfnamefont {F.}~\bibnamefont
  {Villaescusa-Navarro}},\ }\href {https://arxiv.org/abs/1903.03689} {\enquote
  {\bibinfo {title} {Neutrino mass from cosmology: Probing physics beyond the
  standard model},}\ } (\bibinfo {year} {2019}),\ \Eprint
  {http://arxiv.org/abs/1903.03689} {arXiv:1903.03689 [astro-ph.CO]}
  \BibitemShut {NoStop}%
\bibitem [{\citenamefont {{DESI
  Collaboration}}(2024{\natexlab{b}})}]{desi_galaxies}%
  \BibitemOpen
  \bibfield  {author} {\bibinfo {author} {\bibnamefont {{DESI
  Collaboration}}},\ }\href {https://arxiv.org/abs/2404.03000} {} (\bibinfo
  {year} {2024}{\natexlab{b}}),\ \Eprint {http://arxiv.org/abs/2404.03000}
  {arXiv:2404.03000 [astro-ph.CO]} \BibitemShut {NoStop}%
\end{thebibliography}%

\end{document}